\documentclass[12pt]{article}
\usepackage{amsmath,amssymb}

\begin{document}
\providecommand{\V}{\mathcal{V}}
\providecommand{\rh}{\hat{r}}

\begin{titlepage}

\begin{flushright}
                             \texttt{hep-th/0107238}\\
                             DSF--22--2001
\end{flushright}
\vspace{1.5cm}

\begin{center}
\renewcommand{\baselinestretch}{1.8}\normalsize
\textbf{\LARGE On the Phase Transition of Conformal Field Theories
with Holographic Duals}\\[1.5cm]

{\large L.~Cappiello and W.~M{\"{u}}ck}\\[0.5cm]

\renewcommand{\baselinestretch}{1.2}\normalsize
\emph{Dipartimento di Scienze Fisiche, Universit\`{a} di Napoli
``Federico II''}\\
\emph{and}\\
\emph{INFN -- Sezione di Napoli,}\\
\emph{Via Cintia, 80126 Napoli, Italy}\\[12pt]
\emph{E-mails:} \texttt{cappiello,mueck@na.infn.it}\\[1.5cm]
\end{center}

\begin{abstract}
We study the thermodynamic relations of conformal field theories
(CFTs), which are holographically dual to anti-de
Sitter--Schwarzschild bulk space-times.
A Cardy--Verlinde formula is derived thermodynamically for
CFTs living on $S^n{\times} \mathbb{R}$ with $S^n$ having an arbitrary
radius. The Hawking--Page phase transition of the CFT is described
using Landau's theory of phase transitions, and an alternative
derivation of the Cardy--Verlinde formula is presented. The condensate
in the high temperature phase is identified as being composed of
radiational matter.
\end{abstract}
\end{titlepage}

\section{Introduction}
Through the holographic principle \cite{tHooft93,Susskind95}, the
thermodynamics of black holes in anti-de Sitter (AdS) space-times
\cite{Hawking83,Brown94} has found a new application in describing the
high-temperature phase of conformal field theories (CFTs) with holographic
duals \cite{Witten98-2}. Thus, the Hawking--Page phase transition has found
a new interpretation as, \emph{e.g.}, the transition between the confining
and de-confining phases in $\mathcal{N}=4$ super Yang--Mills gauge theory.
Recently, Verlinde \cite{Verlinde00} observed that, in these CFTs, a
Cardy-type entropy formula \cite{Cardy86} holds. Subsequently dubbed the
``Cardy--Verlinde formula,'' its appearance in theories dual to various
bulk space-times has been studied
\cite{Lin00,Nojiri00,Cai01a,Biswas01,Nojiri01a,
Birmingham01,Klemm01b,Youm01a,Nojiri01b,Youm01b,Nojiri01c,Cai01c,Youm01c,Wang01b}.
Black hole phase transitions have recently been discussed in
\cite{Surya01,Stephens01,Creminelli01}.  For a recent list of references on
cosmological aspects of holography and entropy bounds in theories with
holographic duals, see \cite{Youm01c}. See also
\cite{Shiromizu01a,Shiromizu01b} for related topics.

In this letter, we address three questions related to the
thermodynamics of CFTs dual to AdS--Schwarzschild bulk space-times. In
\cite{Verlinde00,Savonije01}, Verlinde and Savonije used Witten's
result \cite{Witten98-2} for the entropy and energy of the boundary
CFT and rescaled the energy and temperature according to the red-shift for
an observer on a brane at a certain distance from the black hole. Although this
is a valid approach, an alternative would be to derive these relations
directly from the on-shell action, which we shall do in Sec.\
\ref{entropy}. Our calculation is based on Witten's \cite{Witten98-2},
but we will find it crucial to express the gravitational
constant in terms of the AdS length scale and a dimensionless
parameter. We analyze the thermodynamics of the AdS--Schwarzschild
black hole and re-derive the Cardy--Verlinde formula.
In Sec.\ \ref{landau}, we shall study the Hawking--Page
phase transition of the boundary CFT using Landau's phenomenological
theory of first order phase transitions \cite{Landau9}. The
thermodynamical relations for energy and entropy are confirmed, and an
alternative derivation of the Cardy--Verlinde formula is presented.
Finally, in Sec.\ \ref{onepoint}, we shall study from a
holographic point of view the high temperature phase of the boundary
CFT. The condensate, expressed phenomenologically as a non-zero order
parameter, is found to consist of radiational matter. This agrees with the
observation that a brane moving in the AdS--Schwarzschild background
obeys the cosmological FRW equations of a radiation dominated
universe \cite{Gubser99}.

\section{Black Hole Thermodynamics}
\label{entropy}
The metric for the (Euclidean) AdS--Schwarzschild black hole in
$(n+2)$ dimensions is
\begin{equation}
\label{metric}
  ds^2 = f(r) dt^2 + \frac1{f(r)} dr^2 +r^2 d\Omega_n^2~,
\end{equation}
where
\begin{equation}
\label{fdef}
  f(r) = 1+\frac{r^2}{l^2} -\frac{\mu}{r^{n-1}}~,
\end{equation}
and $d\Omega_n^2$ is the usual metric of an unit $n$-sphere, $S^n$.
$l$ is the AdS length scale, whereas $\mu$ is a parameter
controlling the mass of the black hole. The AdS--Schwarzschild
space-time with metric \eqref{metric} possesses an event horizon,
$r_+$, defined by
\begin{equation}
\label{r+def}
  f(r_+)=0~.
\end{equation}
Notice that there is only one solution to eqn.\ \eqref{r+def}, because
$f(r)$ is monotonous.

For large $r$ the metric \eqref{metric} behaves asymptotically as
\begin{equation}
\label{metricasy1}
  ds^2 = \frac{r^2}{l^2} \left( dt^2 + l^2 d\Omega_n^2 \right)~.
\end{equation}
Hence, asymptotically, it is conformally equivalent to $S^n{\times}
\mathbb{R}$ (or $S^n {\times} S^1$, if the time dimension is compact)
with $S^n$ having radius $l$ and with $t$ as time parameter. More generally,
let us introduce a radius $\rho$ and rewrite eqn.\
\eqref{metricasy1} as
\begin{equation}
\label{metricasy2}
  ds^2 = \frac{r^2}{\rho^2} \left( d\tau^2 + \rho^2 d\Omega_n^2
  \right)~,
\end{equation}
where the new time parameter is
\begin{equation}
\label{taudef}
  \tau = \frac{\rho}l t~.
\end{equation}

According to the holographic principle, a theory living on a certain
bulk space-time with a metric, whose asymptotic behaviour is given by
\eqref{metricasy2}, is holographically dual to a conformal field
theory living on $S^n{\times} \mathbb{R}$ \footnote{The 1-sphere $S^1$ is used
only to define the temperature of the system in thermal
equilibrium.} with $\rho$ being the radius of $S^n$ and $\tau$ being
the (Euclidean) time variable of the boundary theory.

Following Witten \cite{Witten98-2} we shall now derive the
thermodynamic properties of a conformal field theory that is
holographically dual to the AdS--Schwarzschild bulk. The important
entity to be calculated is the value of the action of the bulk theory
for the metric \eqref{metric}. The bulk gravity action is given by
\begin{equation}
\label{action}
\begin{split}
  I &= \frac{\kappa}{l^n} \int d^{n+2}x \sqrt{\tilde{g}} \left[
  -\tilde{R} - \frac{n(n+1)}{l^2} \right] \\
   &\quad + \frac{\kappa}{l^n} \int d^{n+1} x \sqrt{g} \left[ 2H
  +\frac{2n}{l} + \frac{l}{n-1} R + \cdots \right]~,
\end{split}
\end{equation}
where bulk entities are labelled by a tilde, and those unadorned
belong to the cut-off hypersurface inserted at $r=r_\ast$. The second
line in eqn.\ \eqref{action} comprises the Gibbons-Hawking term with
the trace of the extrinsic curvature, $H=H^i_i$, and the remaining terms are
the usual holographic counter terms added in order to render $I$ finite in
the limit $r_\ast\to \infty$. Adding the surface terms is an
alternative to Witten's subtraction of the pure AdS value of the bulk
integral.
Notice also that we have expressed the
gravitational constant using a dimensionless parameter, $\kappa$. This
is a crucial detail that will allow us later to derive the
thermodynamic relations of the boundary CFT for arbitrary $\rho$.

For the metric \eqref{metric}, the
curvature scalar is $\tilde{R} = -(n+1)(n+2)/l^2$, and the intrinsic
and extrinsic curvatures of $r=\mathrm{const.}$ hypersurfaces are
$R=n(n-1)/r^2$ and $H=-\partial_r \sqrt{f(r)} -n \sqrt{f(r)}
/r$, respectively. For finite temperature, the time variable $t$ is
compactified with periodicity $\beta_0$. Moreover, the black hole
space-time is bounded by the event horizon at $r=r_+$ and by the
cut-off boundary at $r=r_\ast$. Hence, eqn.\ \eqref{action} becomes
\begin{equation}
\label{action2}
\begin{split}
  I &= \frac{\kappa \V_n}{l^n} \beta_0 \left[ \frac2{l^2} (r_\ast^{n+1}
  - r_+^{n+1} ) -r_\ast^n \partial_{r_\ast} f(r_\ast) -2n r_\ast^{n-1}
  f(r_\ast) \right. \\
  &\quad \left. +\frac{2n}l r_\ast^n \sqrt{f(r_\ast)} \left( 1+
  \frac{l^2}{2r_\ast^2} \right) + \cdots \right]~.
\end{split}
\end{equation}
Here, $\V_n$ is the volume of the unit $n$-sphere, and the ellipses
denote further counter terms which we have not
written. For large $r_\ast$ we can write approximately
\begin{equation}
\label{fapprox}
  \sqrt{f(r)} \approx \frac{r_\ast}l \left( 1 + \frac{l^2}{2r_\ast^2}
  -\frac{\mu l^2}{2 r_\ast^{n+1}} \right)~,
\end{equation}
so that we obtain
\begin{equation}
\label{action3}
  I = \frac{\kappa \V_n}{l^n} \beta_0 \left( -
  \frac{2}{l^2} r_+^{n+1} + \mu \right)~.
\end{equation}
The terms with positive powers of $r_\ast$ have exactly
cancelled (using also the counter terms that are not written), the
negative powers vanish in the limit $r_\ast \to
\infty$, and the terms with $(r_\ast)^0$ contribute the $\mu$ term.

For non-zero $\mu$, the period $\beta_0$ is uniquely determined by
demanding completeness and smoothness of the metric at the event
horizon (absence of conical singularities),
\begin{equation}
\label{beta0}
  \beta_0 = \frac{4\pi l^2 r_+}{(n+1)r_+^2 +(n-1) l^2}~.
\end{equation}
Then, expressing $\mu$ in terms of $r_+$ using eqn.\ \eqref{r+def}, we
find from eqn.\ \eqref{action3}
\begin{equation}
\label{I}
  I= \frac{4\pi \kappa \V_n}{(n+1) \rh^2 +(n-1)} \rh^n (1-\rh^2)~,
\end{equation}
where we have introduced the dimensionless variable $\rh= r_+/l$.

The on-shell action $I$ is identified with $\beta F$, where
\begin{equation}
\label{betadef}
  \beta = \frac{\rho}l \beta_0 =
  \frac{4\pi\rho\rh}{(n+1)\rh^2 +(n-1)}
\end{equation}
is the inverse temperature measured in the dual boundary field
theory, and $F$ is the free energy.
Hence,
\begin{equation}
\label{F}
  F = \frac{\kappa \V_n}{\rho} \rh^{n-1} (1-\rh^2)~.
\end{equation}
In the boundary field theory, the volume of the thermodynamic system
in equilibrium is $V=\V_n \rho^n$. Thus, we can consider $F$
naturally as a function of $\beta$ and $V$ and apply the whole text
book apparatus of thermodynamics.

First, the energy is
\begin{equation}
\label{E}
  E =  \left.\frac{\partial (\beta F)}{\partial \beta}\right|_V =
  \frac{n \kappa \V_n}{\rho} \rh^{n-1} (1+\rh^2)~.
\end{equation}
Then, the entropy is obtained by
\begin{equation}
\label{S}
  S = \beta(E - F) = 4\pi \kappa \V_n \rh^n~.
\end{equation}
$S$ is proportional to the area of the event horizon measured in units
of the AdS length scale, $l$. Moreover, it is independent of the
volume $V$. Next, the pressure can be found by evaluating
\begin{equation}
\label{p}
  p =  \left.-\frac{\partial F}{\partial V}\right|_\beta =
  \frac{\kappa}{\rho^{n+1}} \rh^{n-1} (1+\rh^2)~,
\end{equation}
so that one can read off the equation of state,
\begin{equation}
\label{eqofstate}
  E = npV~.
\end{equation}
Finally, Gibbs' free energy is
\begin{equation}
\label{G}
  G = F + pV = \frac{2 \kappa \V_n}\rho \rh^{n-1}~.
\end{equation}
According to Verlinde \cite{Verlinde00}, one can define the Casimir
energy by $E_c= nG$. Then, one easily verifies the validity of the
Cardy--Verlinde formula,
\begin{equation}
\label{Cardy}
  S = \frac{2\pi\rho}n \sqrt{E_c(2E-E_c)}~.
\end{equation}
Notice that we have not used here Verlinde's argument splitting the
energy into an extensive and sub-extensive part. However, expressing
in eqn.\ \eqref{E} $\rho$ and $\rh$ by $V$ and $S$, respectively, the
split becomes apparent.

The interpretations arising from using the thermodynamic analogue are very
suggestive. In standard thermodynamics, $\Delta G \ge0$ holds for irreversible
processes. Thus, since $G\sim E_c\sim c$, where $c$ is a generalized
central charge, $\Delta G\ge0$ represents a ``thermodynamic''
Zamolodchikov theorem, and the renormalization group flow is dual to
an irreversible thermodynamic process.

\section{Phase Transition}
\label{landau}
The boundary field theory on $S^n{\times} \mathbb{R}$ exhibits the
Hawking--Page phase transition \cite{Hawking83}. This is understood as
follows. On the one hand, for the pure AdS bulk geometry, the free
energy is identically zero, $F_{AdS}=0$. Moreover, the value of the inverse
temperature $\beta$ is not restricted by eqn.\ \eqref{betadef}, so
that the pure AdS bulk can exist for any temperature. On the
other hand, for the AdS--Schwarzschild geometry, $\beta$ obeys eqn.\
\eqref{betadef}. According to eqn.\ \eqref{betadef}, for fixed $\rho$
(fixed volume), $\beta$ has a maximum for $\rh=\sqrt{(n-1)/(n+1)}$ with
a value $\beta_{max} = 2\pi \rho/\sqrt{n^2-1}$, corresponding to
Verlinde's minimum temperature of the early universe
\cite{Verlinde00}. Thus, for small
temperatures, $T<1/\beta_{max}$, only the pure AdS bulk geometry is
available as a dual description. For $\sqrt{(n-1)/(n+1)}\le \rh <1$,
both, pure AdS and black hole bulks are available, but the
system prefers the pure AdS bulk, since $F_{BH}>F_{AdS}=0$. For
$\rh>1$, corresponding to
\begin{equation}
\label{betacrit}
  \beta < \beta_c = \frac{2\pi\rho}n~,
\end{equation}
$F_{BH}<F_{AdS}=0$, so that the black hole configuration is
preferred. Notice that for $\beta<\beta_{max}$, there are two
solutions $\rh$. However, the solution $\rh <
\sqrt{(n-1)/(n+1)}$ corresponds to the unstable black hole
\cite{Hawking83} and entails a larger free energy.

Interestingly, at the transition point, the energy and
Casimir energy coincide, $E=E_c$. Moreover, eqn.\ \eqref{betacrit}
suggests to rewrite the Cardy--Verlinde formula \eqref{Cardy} as
\begin{equation}
\label{Cardy2}
  S = \beta_c \sqrt{E_c(2E-E_c)}~.
\end{equation}

We shall now try to understand the first order phase transition at
$\beta=\beta_c$ using Landau's phenomenological theory
\cite{Landau9}. We shall choose $\rh$
as the order parameter, with $\rh=0$ for $T<T_c$, and $\rh$ given as
the larger solution of eqn.\ \eqref{betadef} for $T>T_c$. Let us
forget the thermodynamic analysis of the last section
and assume that we only have eqns.\ \eqref{betadef} and \eqref{F}, which
follow from the bulk analysis. The parameter $\rho$ shall be kept
fixed, \emph{i.e.}, we consider a fixed volume.
Eqn.\ \eqref{F} should be interpreted as the free energy at
equilibrium, and we can try to write down a more general expression,
$F(\rh,T)$, describing the system also away from
equilibrium. Analogously, eqn.\ \eqref{betadef} is to be interpreted
as the equilibrium condition relating the temperature and the order
parameter. Eqn.\ \eqref{betadef} can be rewritten in terms of the
transition temperature, $T_c=n/(2\pi\rho)$, as
\begin{equation}
\label{Tdef}
  T = \frac{T_c}{2n\rh} \left[(n+1) \rh^2 +(n-1) \right]~.
\end{equation}
Let us make the ansatz
\begin{equation}
\label{Fansatz}
  F(\rh,T) = \frac{\kappa \V_n}{\rho} \left( a \rh^{n-1} - bT \rh^n +c
  \rh^{n+1} \right)~,
\end{equation}
where $a$, $b$ and $c$ are three constants to be
determined. Substituting into eqn.\ \eqref{Fansatz} the equilibrium
temperature, eqn.\ \eqref{Tdef}, and comparing with
the equilibrium expression for $F$, eqn.\ \eqref{F}, we find
\begin{equation}
\label{ac}
\begin{split}
  a &= \frac{(n-1)bT_c}{2n}+1~,\\
  c &= \frac{(n+1)bT_c}{2n}-1~.
\end{split}
\end{equation}
In addition, we demand that the equilibrium
condition derived from eqn.\ \eqref{Fansatz},
\begin{equation}
\label{eqcond}
  \frac{\partial F(\rh,T)}{\partial \rh} =0~,
\end{equation}
coincide with eqn.\ \eqref{Tdef}. This yields
\begin{equation}
\label{eqcond2}
  T = \frac{1}{nb\rh} \left[c(n+1) \rh^2 +a(n-1) \right]~,
\end{equation}
which, together with eqn.\ \eqref{ac}, leads to
\begin{equation}
\label{abc}
  a=c=n~, \quad  b=2n \beta_c~.
\end{equation}

Thus, we have the free energy from eqn.\ \eqref{Fansatz},
\begin{equation}
\label{Free}
  F(\rh,T) = \frac{\kappa\V_n n}{\rho} \left( \rh^{n-1} -2 \beta_c T
  \rh^n + \rh^{n+1} \right)~.
\end{equation}
The free energy \eqref{Free} correctly describes the first order phase
transition at $T=T_c$ in terms of the order parameter $\rh$. In fact,
for fixed $T<T_c$, the absolute minimum of $F$ is at $\rh=0$, whereas
for $T>T_c$, $F$ assumes its absolute minimum for $\rh$ given by the
larger solution of eqn.\ \eqref{betadef}. The smaller solution
corresponds to the local maximum of $F$.

In equilibrium, we have the entropy
\begin{equation}
\label{Seq}
  S = - \frac{dF(\rh,T)}{dT} = - \frac{\partial F(\rh,T)}{\partial T}
  = \frac {2n \kappa \V_n}{\rho} \beta_c \rh^n = 4\pi\kappa \V_n \rh^n~,
\end{equation}
where we have made use of eqn.\ \eqref{eqcond}. Eqn.\ \eqref{Seq} is
in agreement with eqn.\ \eqref{S}. We also confirm the expression
\eqref{E} for the energy at equilibrium,
\begin{equation}
\label{Eeq}
  E = F + TS = \frac{n \kappa \V_n}{\rho} \rh^{n-1} (1+\rh^2)~.
\end{equation}

We shall now give another, more general, derivation of the
Cardy--Verlinde formula. Let us rewrite eqn.\ \eqref{Free}
generically as
\begin{equation}
\label{Free2}
  F(\rh,T) = \frac12 E_c(\rh) \left( 1 -2 \beta_c T \rh + \rh^2
  \right)~,
\end{equation}
where we have introduced an energy $E_c(\rh)$ that depends only on the
order parameter. In eqn.\ \eqref{Free}, $E_c$ coincides with the
thermodynamic Casimir energy of the CFT dual to the
AdS--Schwarzschild bulk. In general,
we shall assume that $E_c(\rh)$ grows monotonously, and that
$E_c(0)=0$.  Then, $F(\rh,T)$ describes a system with a first order phase
transition from $\rh=0$ for $T<T_c$ to some $\rh(T)>0$ for $T>T_c$.
We find the equilibrium entropy and energy,
\begin{align}
\label{Seq2}
  S &= \beta_c E_c(\rh) \rh~,\\
\label{Eeq2}
  E &= \frac12 E_c(\rh) (1+\rh^2)~.
\end{align}
Obviously,
\begin{equation}
\label{Cardy3}
  E_c (2E-E_c) = (E_c \rh)^2 = (ST_c)^2~,
\end{equation}
which is the Cardy--Verlinde formula in the form \eqref{Cardy2}.
Thus, the Cardy--Verlinde formula arises generically in theories with
first order phase transitions, which are described by a free energy
of the form \eqref{Free2}.

Finally, let us try to give generic interpretations of
$E_c(\rh)$. First, from eqn.\ \eqref{Eeq2} follows that
$E=E_c$ for $T\le T_c$ ($E=E_c=0$ for $T<T_c$), whereas $E>E_c$ for
$T>T_c$. Second, in a microcanonical ensemble, we should
consider $S$ and $E_c$  as a function of the energy and possibly other
parameters, call them $x$. Then, from eqn.\ \eqref{Cardy3} follows
\begin{equation}
\label{Ec_interpret}
  T_c^2 S \left.\frac{\partial S}{\partial x}\right|_E
  = (E-E_c) \left.\frac{\partial E_c}{\partial x}\right|_E~,
\end{equation}
so that $E_c$ is the energy of the microcanonical ensemble at which the
entropy of the system is stationary under changes of the other system
parameters.

\section{Holographic Description of the Condensate}
\label{onepoint}
In the previous section, we have discussed from a thermodynamic point
of view the phase transition of the boundary CFT. The theory acquires
a non-zero order parameter in the high temperature phase.
We would like to understand this process also from a holographic point
of view. The non-zero order parameter should manifest itself as a
physical condensate of some quantum field.
For this purpose, let us consider the one-point function of
the energy momentum tensor of the boundary CFT.
It is obtained from the first order variation of the action $I$,
eqn.\ \eqref{action}, under the variation of the metric,
$g_{ij}\to g_{ik}(\delta^k_j +h^k_j)$. One finds
\begin{equation}
\label{dI}
  \delta I = \frac{\kappa}{l^n} \int d^{n+1} x \sqrt{g} \left[ -H^i_j
  +H \delta^i_j +\frac{n}{l} \delta^i_j + \frac{l}{n-1} \left( -R^i_j
  + \frac12 R \delta^i_j \right) + \cdots \right] h^j_i~,
\end{equation}
where the bulk term vanishes, because we are on-shell.
The integral is evaluated at the cut-off, $r=r_\ast$, and the counter
terms ensure that $\delta I$ is finite in the limit $r_\ast\to
\infty$. In the course of the calculation one realizes that the
counter terms, which are not written explicitly, do not contribute to
the finite result, but serve only to remove the divergent terms.

With
\begin{equation}
\label{HRrels}
\begin{aligned}
  H^t_t &= - \partial_r \sqrt{f(r)}~,&
  H^\alpha_\beta &= -\frac1r \sqrt{f(r)} \delta^\alpha_\beta~,\\
  R_{tt} &= R_{t\alpha}= H_{t\alpha}=0~,&
  R^\alpha_\beta &= \frac{n-1}r \delta^\alpha_\beta~,
\end{aligned}
\end{equation}
($\alpha, \beta=1,\ldots,n$; $t$ as index denotes the $\mathbf{e}_t$-components
of the tensors) and using eqns.\ \eqref{fdef} and \eqref{fapprox}, we find
\begin{equation}
\label{dI2}
  \delta I = \frac{\kappa}{2l^n} \int d^{n+1}x \left( n\mu h^t_t - \mu
  \delta^\alpha_\beta h^\beta_\alpha \right) + \text{anomaly terms}~.
\end{equation}
The anomaly terms arise from additional counter terms to be added when
$n$ is odd \cite{Henningson98-2}.
They contribute a gravitational anomaly to the energy
momentum tensor one-point function, but we shall not consider them
explicitly. Thus, from
\begin{equation}
\label{thetadef}
 \left\langle \Theta^i_j\right\rangle = -2 \frac{\delta I}{\delta h^j_i}~
\end{equation}
and using eqn.\ \eqref{r+def} to eliminate $\mu$, we find the energy
momentum tensor one-point functions,
\begin{equation}
\label{theta}
  \left\langle\Theta^t_t\right\rangle
  = - \frac{n\kappa}l \rh^{n-1} (1+\rh^2)~,\quad
  \left\langle\Theta^\alpha_\beta\right\rangle
  = \frac{\kappa}l \rh^{n-1} (1+\rh^2) \delta^\alpha_\beta~,
\end{equation}
apart from the possible gravitational anomaly. $\Theta^i_j$ is the
energy momentum tensor of the boundary CFT living on $S^n{\times}
\mathbb {R}$, with $t$ as time parameter and $l$ being the radius of
$S^n$. In order to obtain the energy momentum tensor for an arbitrary
radius $\rho$, one simply multiplies the expressions in eqn.\
\eqref{theta} by $l/\rho$. This yields
\begin{equation}
\label{theta2}
  \left\langle\Theta^\tau_\tau \right\rangle
  = - \frac{E}{\V_n}~,\quad
  \left\langle\Theta^\alpha_\beta\right\rangle
  = \rho^n p \delta^\alpha_\beta~,
\end{equation}
with $E$ and $p$ given by eqns.\ \eqref{E} and \eqref{p}, respectively.
Obviously, $\Theta^i_i=0$ (up to the gravitational anomaly), which
corresponds to radiation-type matter. Thus, we conclude that the boundary
CFT contains a (temperature dependent) condensate of radiational matter in
the high temperature phase. From a bulk point of view, this radiation
consists of the Hawking radiation emitted from the black hole and
traversing the brane of the observer.

\section{Conclusions}
In this paper we have studied the thermodynamics of CFTs which are
holographically dual to AdS-Schwarzschild black holes. We have analyzed the
Hawking-Page phase transition using  Landau's theory of first order phase
transitions. This has led us to an alternative derivation of the
Cardy-Verlinde formula. In the course of our analysis we have reformulated
some relations for the thermodynamical quantities in a way which is
less dependent on the specific form of the AdS black hole solution,
suggesting a broader range of validity of the  Cardy-Verlinde
formula. This is especially true in the expression \eqref{Free2} for
the free energy, which allows to derive the Cardy-Verlinde formula
under the simple hypothesis of monotonicity of the function
$E_c(\rh)$, interpreted as the order parameter dependent Casimir
energy. It might be interesting to find statistical systems which show
the generic behaviour leading to the Cardy-Verlinde formula, providing
microscopic interpretations of the Casimir energy.

We have also used the bulk action to show that, at the classical
level, the high temperature phase of the holographically dual matter
satisfies the equation of state of radiation, in agreement with the recent
construction of FRW space-times on branes moving in a AdS-Schwarzschild
background.

\section*{Acknowledgments}
W.~M.\ gratefully acknowledges financial support by the European
Commission RTN programme HPRN-CT-2000-00131, in which
he is associated with INFN, Sezione di Frascati.

\end{document}